# Effective resistivity for magnetohydrodynamic simulation of collisionless magnetic reconnection

**H. W. Zhang[1, 2, *], Z. W. Ma[2, *], and T. Chen[2]**

[1]Max Planck Institute for Plasma Physics, Boltzmannstr. 2, 85748 Garching b. M., Germany

[2]Institute for Fusion Theory and Simulation, School of Physics, Zhejiang University, 310027, Hangzhou, China

[*]Corresponding authors: H. W. Zhang (haowei.zhang@ipp.mpg.de); Z. W. Ma (zwma@zju.edu.cn)

**Key points:**

- We derive an effective resistivity model for collisionless magnetic reconnection and apply it to magnetohydrodynamic simulations
- The effective resistivity improves the reconnection rate to the order of $0.1B_0v_A$
- The properties of the simulated current sheets are quite consistent with the particle-in-cell simulations

**Abstract**

The electron inertia and the off-diagonal electron pressure terms are well-known for the frozen-in condition breakdown in collisionless magnetic reconnection, which are naturally kinetic and difficult to employ in magnetohydrodynamic (MHD) simulations. Considering the limitations of MHD and Hall MHD in neglecting the important electron dynamics such as the inertia and the nongyrotropic pressure, the kinetic characteristics of electrons and ions in the diffusion region are studied, and an effective resistivity model involving dynamics of charged particles is proposed *(Ma et al. 2018 Sci. Rep. 8 10521)*. The amplitude of the effective resistivity is mainly determined by electrons in most realistic situations with large ion-electron mass ratios. In this work, the effective resistivity model for collisionless magnetic reconnection without the guide field is successfully applied in the 2.5D MHD and Hall MHD simulations, which remarkably improves the simulation results compared with traditional MHD models. For the MHD case, the effective resistivity significantly increases the reconnection rate to a reasonable value of $\sim 0.1 B_0 v_A$. For the Hall MHD case with effective resistivity, the peak reconnection rate is $\sim 0.25 B_0 v_A$, and the major structures of the reconnecting field and the current sheet agree well with the particle-in-cell (PIC) and hybrid simulations.

**Plain Language Summary**

Magnetic reconnection is a fundamental process in space and laboratory plasmas. Despite the collisionless nature of many of these systems, magnetic reconnections occur rapidly, thereby indicating the presence of an effective or anomalous resistivity in the reconnection region. A lot of theoretical and experimental research has been conducted to explain the mechanism of collisionless reconnection, and a consensus has emerged regarding the key role of charged particle dynamics in the reconnection region. In this study, we quantitatively estimate the effective resistivity based on the kinetic behaviour of particles and successfully implement it in magnetohydrodynamic (MHD) simulations, which are much more efficient than particle-in-cell (PIC) simulations. The effective resistivity model replicates the reasonable reconnection rate and improves the current sheet structure in MHD simulations, closely matching PIC simulations. This work bridges the gap between subtle small-scale dynamics and practical large-scale models, and contributes to the development of extended MHD models for more efficient and accurate study of collisionless plasmas.

## 1 Introduction

Magnetic reconnection, characterized by energy conversion and transport processes, plays an important role in the topological evolution of magnetized plasmas in both space and laboratory systems. The concept of magnetic reconnection was first suggested by *Giovanelli* (1946), and the first well-known model was proposed by *Sweet* (1958) and *Parker* (1957). However, the predicted reconnection rate with the Y-type geometry of the Sweet-Parker model is too low to explain explosive phenomena, such as solar flares, magnetospheric substorms (*Eugene Newman Parker*, 1979), and tokamak disruptions (*Taylor*, 1986). In contrast, the Petschek model predicts a much faster reconnection rate by considering the X-type structure in a smaller diffusion region (*Petschek*, 1964). Though the Petschek-type configuration has been confirmed in various simulations by including, for example, locally enhanced resistivity (*Ugai*, 1995) and Hall effect (*Ma et al.*, 2015), a critical issue is that such a tiny structure can hardly form in most high-S collisionless plasma simulations.

Resistivity or equivalent magnetic diffusion mechanism is critical for breaking the frozen-in condition and triggering magnetic reconnection. For example, the impact of the localized resistivity with different magnitudes and profiles on the reconnection was studied via MHD simulations (*Jiménez et al.*, 2022), which predicts a highest normalized reconnection rate of approximately 0.25 . However, in collisionless plasmas, the Spitzer resistivity (*Spitzer*, 2006) based upon electron-ion collision is too small to explain the fast magnetic reconnection (*Speiser*, 1970). A number of studies have been carried out to investigate the anomalous resistivity in collisionless magnetic reconnection. The effective conductivity determined by the inertia and the gyromotion of the particles rather than the particle-particle collisions or wave-particle collisions, i.e., the lifetime of particles in the diffusion region and the gyro period outside the diffusion region, was first studied for the current sheet in the geomagnetic tail (*Speiser*, 1970). Fast reconnection was obtained in simulations by setting resistivity as functions of relative electron-ion drift velocity (*Ugai*, 1995; *Yokoyama and Shibata*, 1994). The anomalous resistivity model depending on plasma current was studied in magnetohydrodynamic (MHD) and Hall MHD simulations (*Otto*, 2001). The chaos-induced effective resistivity by analyzing the chaotic motion of particles around the X-point was suggested (*Numata and Yoshida*, 2002). The off-diagonal plasma pressure tensor terms were found to be responsible for relaxing the frozen-in condition (*Drake and Shay*, 2007) with particle-in-cell (PIC) simulations (*Cai and Lee*, 1997; *Pritchett*, 2001) and hybrid simulations (*Kuznetsova et al.*, 2001). The critical role of electron inertia in Hall MHD simulations of collisionless reconnection was confirmed (*Andrés et al.*, 2014). Recently, a kinetic physics-motivated effective resistivity model based on the full Ohm's law derived from first principles but expressed by fluid quantities was proposed and implemented in 2D resistive relativistic MHD simulations for electron-positron pair plasma (*Bugli et al.*, 2025; *Selvi et al.*, 2023) and it was further improved by the empirical prescription based on PIC simulations (*Moran et al.*, 2025). Besides the simulation efforts, the Sweet-Parker model was refined by incorporating the compressibility, the downstream pressure, and the effective resistivity to explain the results of the Magnetic Reconnection Experiment (MRX) (*Ji et al.*, 1999).

Consensus has been reached on the significance of electron dynamics in frozen-in condition breakdown within the diffusion region (*Drake and Shay*, 2007). The limitations of MHD and Hall MHD models in neglecting the electron inertia term and nongyrotropic pressure, result in challenges in describing small-scale kinetic effects in the diffusion region. In this context, our previous work theoretically studied the kinetic mechanism of effective or anomalous resistivity in collisionless magnetic reconnection based on characteristic motions of electrons and ions in the diffusion region, and an effective resistivity model has been suggested (*Ma et al.*, 2018). The mechanism of the effective resistivity is mainly determined by electron dynamics in most realistic cases with large ion-electron mass ratios. The estimated effective resistivity has been compared against the values from PIC simulations, which shows quantitative agreement.

In this work, the proposed effective resistivity model is successfully applied in 2.5D MHD and Hall MHD simulations without the guide magnetic field. With the effective resistivity, the reconnection rate, the topologies of the reconnecting field and the current sheet are significantly improved in both MHD and Hall MHD simulations. Specifically, the MHD simulation with effective resistivity predicts the reconnection rate at a reasonable level of $0.1\ B_0 v_A$ (*Comisso and Bhattacharjee*, 2016). The Hall MHD simulation results with effective resistivity are much more consistent with the PIC and hybrid simulation results, with a reconnection rate of $\sim 0.25 B_0 v_A$. The results further demonstrate the importance of electron dynamics in the diffusion region.

The remainder of this paper is organized as follows. In Section 2, the effective resistivity model (*Ma et al.*, 2018) is briefly reviewed. The extended MHD model, including the Hall term and the effective resistivity, is introduced in Section 3. Section 4 presents the comparisons of the MHD simulations with and without effective resistivity. Similarly, Section 5 compares the Hall MHD simulation cases with and without effective resistivity. The summary and discussion are presented in Section 6.

## 2 The effective resistivity model

For magnetic reconnection in 2.5D slab geometry without the guide field (Harris equilibrium), the out-of-plane flow (in the $y$ direction) of charged particles determines the topology and intensity of the central current sheet. Accordingly, the main idea of *Ma et al.* (2018) is to analyze the characteristic motion of charged particles in the reconnecting field. As shown in **Figure 2** (or the schematic by Figure 1 in *(Ma et al., 2018)*), the bulk velocity and current of plasma in the diffusion region are mainly in the out-of-plane direction. The Lorentz force by the bending magnetic field tends to change the motion direction of charged particles downstream ($x$ direction), which is equivalent to scattering the particles away from the diffusion region and preventing the particle from being continuously accelerated by the out-of-plane reconnecting electric field. As a result, the bending magnetic field induced pitch-angle scattering determines a characteristic timescale on the electric-particle acceleration. The statistical effect for all particles is equivalent to the enhancement of out-of-plane resistivity.

To estimate the effective resistivity induced by pitch-angle scattering, we investigate the kinetic motions of charged particles in electromagnetic field around the X-point. Without loss of generality, we first consider the electron, the motion equation of electrons due to electromagnetic force is

$$\frac{d\mathbf{v}_e}{dt} = -\frac{e}{m_e}\left(\frac{\mathbf{v}_e \times \mathbf{B}}{c} + \mathbf{E}\right). \tag{1}$$

Perform the first-order expansions for the magnetic field around an arbitrary point $(x_0, z_0)$ near the X-point in the $x$-$z$ plane (the initial magnetic field is in the $x$ direction, and the magnetic field strength changes in the $z$ direction)

$$\begin{aligned}\mathbf{B} = B_x\left(1 + \frac{z - z_0}{L_{xz}} + \frac{x - x_0}{L_{xx}}\right)\hat{\boldsymbol{x}} \\ +B_z\left(1 + \frac{x - x_0}{L_{zx}} + \frac{z - z_0}{L_{zz}}\right)\hat{\boldsymbol{z}} + B_y\hat{\boldsymbol{y}},\end{aligned} \tag{2}$$

$$L_{ij} = B_i/\partial_j B_i, (i, j = x, z), \tag{3}$$

where $L_{ij}$ is the characteristic length for $B_i$ in the $j$ direction.

To simplify the derivation, several assumptions for the diffusion region are given. First, the dominant reconnecting electric field is out-of-plane $(E_y)$, and the shear magnetic field is in-plane (without the guide magnetic field). Therefore, the in-plane electric field and out-of-plane magnetic field parts are omitted in Eq. (1). Second, the evolutionary timescale for the central current is much longer than the characteristic timescale of pitch-angle scattering. Then, we can estimate the averaged out-of-plane electron speed by $e\bar{v}_{ey}/m_e = J_{ey}/\rho_e$ and replace $v_{ey}$ by $\bar{v}_{ey}$ in Eq. (1). Thirdly, due to the directional nature of Lorentz force, the terms with $L_{xz}$, $L_{zx}$ outweigh those with $L_{xx}$, $L_{zz}$ much more. Finally, the sheared $B_x$ leads to a quasi-oscillation in the $z$ direction, while $B_z$ results in the electron scattering in the $x$ direction, the two processes being almost independent of each other. Thus, the effective resistivity by pitch-angle scattering is mainly contributed by the reconnecting field component $B_z$. Combing Eqs. (1) - (2) and the above assumptions, the electron motion Eq. (1) can be reduced into the $x$ component

$$\frac{d^2x}{dt^2} = \frac{J_{ey}B_z}{\rho_e L_{zx} c}(L_{zx} + x - x_0). \tag{4}$$

An effective time scale for electric acceleration $\tau_e$ is defined as the duration that the electron spends on leaving the $L_{zx}$ downstream away from the initial point by the pitch-angle scattering process. We obtain the analytical solution $x(t)$ and $\tau_e$ for $|x(\tau_e) - x_0| = L_{zx}$ (*Ma et al.*, 2018). Through further approximation by ignoring $v_{x0}$ around the X-point, $\tau_e$ is represented by

$$\tau_e \approx \sqrt{\rho_e L_{zx} c / J_{ey} B_z}. \tag{5}$$

The out-of-plane variation tendencies of electron velocity $v_{ey}$ and the current density $J_{ey}$ around the X-point due to electric field $E_y$ during $\tau_e$ are

$$\delta v_{ey} = e \tau_e E_y / m_e, \tag{6}$$

$$\delta J_{ey} = n_e e^2 \tau_e E_y / m_e. \tag{7}$$

Eq. (7) indicates the effective resistivity around the X-point for $J_{ey}$ in $y$ direction as

$$\eta_e = m_e^2 / e^2 \rho_e \tau_e. \tag{8}$$

The similar effective resistivity for ions can be derived by considering the ion current density $J_{iy}$. Besides, the electric field $E_y$ equals to the products of effective resistivity $\eta_s$ (neglecting the collisional Spitzer resistivity based on collisionless assumption) and current density $J_{sy}$ for each species (character $s$ indicating ion and electron), the electron and ion effective resistivities satisfy

$$\frac{\eta_e}{\eta_i} = \frac{J_{iy}}{J_{ey}} \approx \sqrt{\frac{m_e J_{ey}}{m_i J_{iy}}}. \tag{9}$$

The total effective resistivity $\eta_{eff}$ around the X-point is

$$\eta_{eff} \approx \frac{1}{1 + \sqrt{m_e J_{ey} / m_i J_{iy}}} \eta_e \approx \frac{m_e^2 / e^2 \rho_e}{1 + \sqrt{m_e J_{ey} / m_i J_{iy}}} \sqrt{\frac{J_{ey} B_z}{\rho_e L_{zx} c}}. \tag{10}$$

With $m_i / m_e \approx 1836$ and combining Eqs. (9) - (10), we have

$$\eta_{eff} \approx 0.9 \eta_e. \tag{11}$$

Eq. (11) demonstrates that in most realistic situations with a high ion-electron mass ratio, the electron dynamics plays the leading role in the total effective resistivity. According to Eq. (10), the effective resistivity strongly depends on the spatial characteristic of reconnecting magnetic field around the X-point $(B_z / L_{zx})$. When magnetic reconnection occurs, $B_z$ increases and $L_{zx}$ decreases, leading to the enhancement of the effective resistivity. Detailed derivation and the quantitative verification for the effective resistivity by PIC simulations are reported in (*Ma et al.*, 2018).

## 3 Extended MHD simulation model and the initial equilibrium

The compressible 2.5D (uniform in the $y$ direction, i.e., $\partial / \partial y = 0$) extended MHD model, including the Hall effect and the effective resistivity, is employed. The simulations are performed in the Cartesian coordinate system within a rectangular box of $-L_x \le x \le L_x$, $-L_z \le z \le L_z$. The magnetic field is represented by the magnetic flux $\psi(x,z,t)$

$$\mathbf{B} = \hat{\boldsymbol{y}} \times \nabla\psi(x,z,t) + B_y(x,z,t)\hat{\boldsymbol{y}}. \tag{12}$$

The compressible Hall MHD equations are *(Ma and Bhattacharjee, 2001*)

$$\frac{\partial\rho}{\partial t} = -\nabla\cdot(\rho\mathbf{v}), \tag{13}$$

$$\frac{\partial(\rho\mathbf{v})}{\partial t} = -\nabla\cdot[\rho\mathbf{v}\mathbf{v} + (p + B^2/2\,)\mathbf{I} - \mathbf{B}\mathbf{B}], \tag{14}$$

$$\frac{\partial\psi}{\partial t} = -\mathbf{v}\cdot\nabla\psi + \frac{1}{S_{tot}}J_y + \frac{d_i}{\rho}(\mathbf{J}\times\mathbf{B})_y, \tag{15}$$

$$\frac{\partial B_y}{\partial t} = -\nabla\cdot\left(B_y\mathbf{v}\right) + \mathbf{B}\cdot\nabla v_y + \frac{1}{S_{spz}}\nabla^2 B_y - d_i\nabla\left[\nabla\times\left(\frac{\mathbf{J}\times\mathbf{B} - \nabla p}{\rho}\right)\right]_y, \tag{16}$$

$$\frac{\partial p}{\partial t} = -\nabla\cdot(p\mathbf{v}) - (\gamma - 1)p\nabla\cdot\mathbf{v} + \frac{1}{S_{tot}}J_y^2 + \frac{1}{S_{spz}}(J_x^2 + J_z^2). \tag{17}$$

where **v**, **B**, **J**, $\psi$, $\rho$, $p$, **I** are plasma velocity, magnetic field, current density, flux function, plasma mass density, thermal pressure, and unit tensor, respectively. All variables are normalized by: $\mathbf{B}/B_0 \to \mathbf{B}$, $\boldsymbol{x}/d_i \to \boldsymbol{x}$, $\mathbf{v}/v_A \to \mathbf{v}$, $t/\tau_A \to t$, $\psi/(B_0 d_i) \to \psi$, $\rho/\rho_0 \to \rho$, and $p/(B_0^2/4\pi) \to p$, where $B_0$ is the initial asymptotic magnetic strength, $d_i$ is the ion inertial length, $v_A = B_0/(4\pi\rho_0)^{1/2}$ is the Alfvén velocity, $\tau_A = d_i/v_A = \omega_{ci}^{-1}$ is the Alfvén time (equivalent to the inverse of the ion cyclotron frequency), $\rho_0$ is the asymptotic mass density, $\gamma = 5/3$ is the ratio of specific heats of plasma. The relative change of $d_i$ from 0 to 1 in Eqs. (15) - (16) represents the intensity of Hall effect included in the simulation.

Due to the anisotropy of effective resistivity, two different Lundquist numbers $S_{tot(spz)} = \tau_{R,tot(spz)}/\tau_A$ are applied out-of-plane $[\tau_{R,tot} = 4\pi d_i^2/c^2\left(\eta_{spz} + \eta_{eff}\right)]$ and in-plane $(\tau_{R,spz} = 4\pi d_i^2/c^2\eta_{spz})$, respectively, where $\eta_{spz}$ and $\eta_{eff}$ are the Spitzer resistivity and effective resistivity, $c$ is the speed of light. For the Spitzer resistivity, a large constant Lundquist number $(S_{spz} = 1000)$ indicating low collisionality is adopted in all simulations. The effective resistivity induced $S_{eff}$ is calculated in the following manner.

As shown by **Figure 2**, for a typical reconnecting magnetic field pattern with the X-point at $(x_0 = 0, z_0 = 0)$, two symmetrical extreme points of $B_z$ can be found in the left and right half panels, marked as $(\pm x_1, z_1 = 0)$. Then, a specified point with the location $(x_2 = x_1/2, z_2 = 0)$ at the halfway from the X-point to the right extreme point of $B_z$ is chosen to estimate the spatial variation of $B_z$ around X-point. The value of $B_z/L_{zx}$ in Eq. (10) is estimated with $\delta B_z/(x_2 - x_0)$, where $\delta B_z$ is the $z$ component of the magnetic field strength at $(x_2, 0)$. Based on the assumption of similar average kinetic energies of electrons and ions, we can estimate the value of $\eta_{eff}$ from Eq. (10). Using the same normalization as above, $S_{eff}(x,z,t)$ at each point can be estimated by

$$\frac{1}{S_{eff}(x,z,t)} \approx \kappa \cdot \sqrt{\left|\frac{J_y \delta B_z}{\rho^3 (x_2 - x_0)}\right|} \cdot \exp\left[-\left(\frac{z - z_0}{\lambda_b}\right)^2 - \left(\frac{x - x_0}{x_2 - x_0}\right)^2\right]. \tag{18}$$

The coefficient $\kappa \approx 0.9\sqrt{m_e/m_i}$ is a constant calculated based on the mass ratio and Eqs. (10) - (11), for example, $\kappa \approx 0.02$ for $m_i/m_e \approx 1836$. The spatial scales of the effective resistivity are the real-time half-width $\lambda_b$ of the current sheet and characteristic length $L_{zx} = x_2 - x_0$ of the reconnecting magnetic field $B_z$.

The initial plasma velocity is zero. The asymptotic plasma beta $\beta = 0.1$. The thermal pressure is obtained by solving the equilibrium equation

$$p = (1 + \beta) B_0^2/2 - B^2/2. \tag{19}$$

A classical Harris current sheet with a half-width of $\lambda_b$ is chosen as the initial state

$$B_x = B_0 \tanh(z/\lambda_b), B_y = B_z = 0. \tag{20}$$

The initial mass density profile is given by

$$\rho(z) = \rho_0 + \rho_1 \operatorname{sech}^2(z/\lambda_\rho). \tag{21}$$

The mass density is set as $\rho_0 = 1.0$ and $\rho_1 = 0.2$. The characteristic scales for magnetic field and mass density are $\lambda_b = \lambda_\rho = 0.5$.

Eqs. (13) - (17) are solved with the fourth-order Runge-Kutta method in time and the fourth-order finite difference method in space. The time step $\Delta t$ is determined by the Courant–Friedrichs–Lewy (CFL) condition. The simulation box is $-12.8 \le L_x \le 12.8$ and $-6.4 \le L_z \le 6.4$, with 640×1000 grid points uniformly distributed in the $x$ direction and nonuniformly distributed in the $z$ direction $(dx = 0.04, dz_{min} = 0.01, dz_{max} = 0.02)$. Periodic boundary condition in the $x$ direction and free boundary condition in the $z$ direction are adopted.

The reconnection rate $\gamma_{rate}$ is calculated by the time derivation of the flux function or the product of resistivity and out-of-plane current density at the X-point, which should be almost equivalent in the absence of numerical diffusion, that is

$$\gamma_{rate} = \partial\psi/\partial t \simeq \eta J_y. \tag{22}$$

The reconnection process is triggered with a small magnetic perturbation $(\delta\psi_0 = 0.01)$

$$\delta\psi = \delta\psi_0 \cos(\pi x/L_x)\cos(\pi z/2L_z). \tag{23}$$

## 4 MHD simulation results

First, we apply the effectivity resistivity model of Eq. (18) in the resistive MHD model without Hall effect $(d_i = 0)$. For the case without effective resistivity ($\eta_{spz}$ case), the coefficient $\kappa$ in Eq. (18) is set to 0 such that only the constant Spitzer resistivity $(1/S_{spz} = 0.001)$ is considered. For the case with the effective resistivity ($\eta_{eff}$ case), $\kappa$ is set to 0.02 based on previous estimation $(m_i/m_e \approx 1836)$.

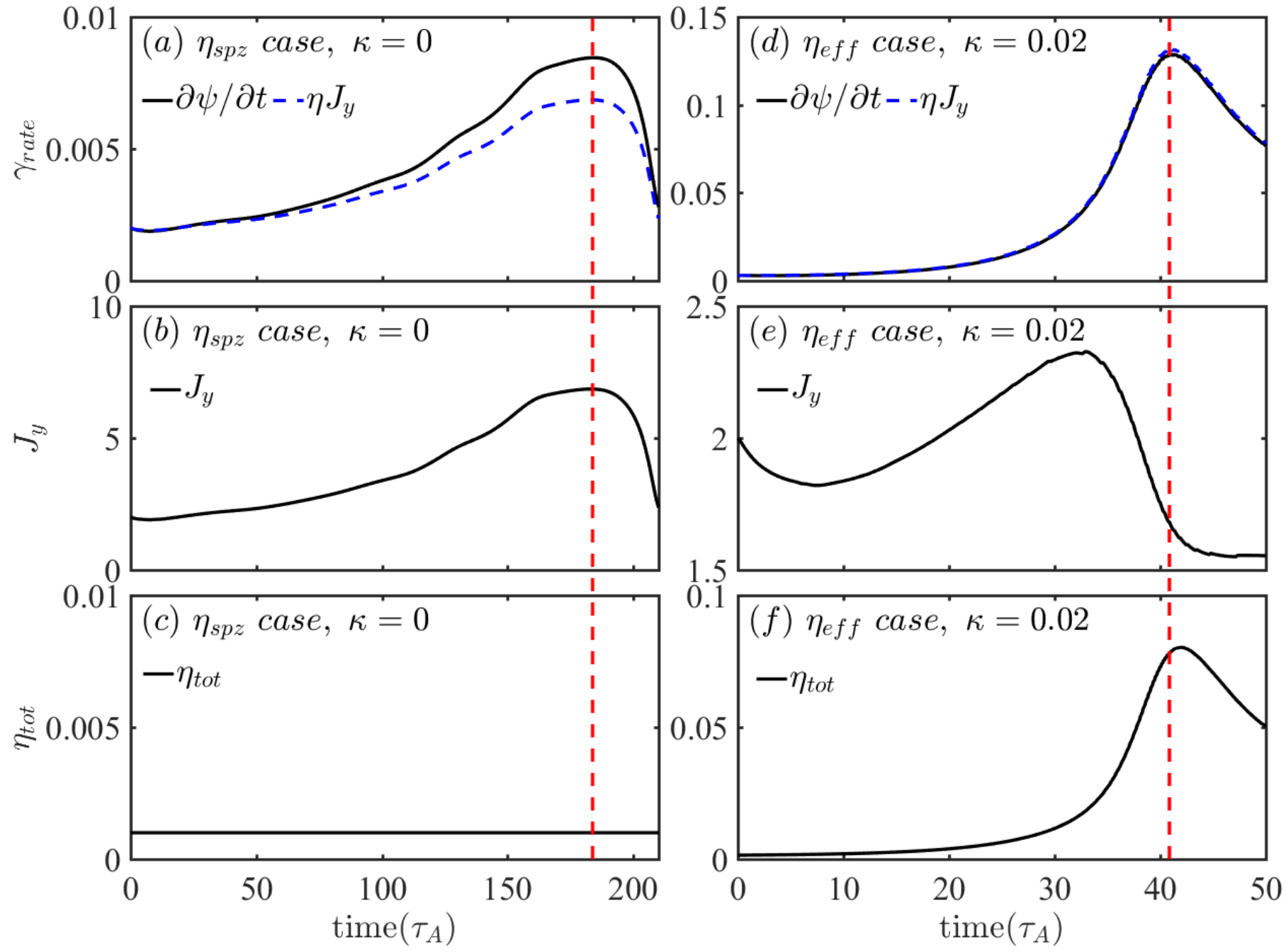

**Figure 1.** (MHD) Time evolution results for (left) the $\eta_{spz}$ case with $\kappa = 0$ and (right) the $\eta_{eff}$ case with $\kappa = 0.02$: (top) reconnection rate calculated by $\partial\psi/\partial t$ (solid line) and $\eta J_y$ (dashed line); (middle) the out-of-plane current density $J_y$ at the X-point; (bottom) the total resistivity $\eta_{tot}$. The dashed lines mark the moments of peak reconnection, $t_{peak} = 180\tau_A$ for $\eta_{spz}$ case and $t_{peak} = 41\tau_A$ for $\eta_{eff}$ case, respectively.

The time evolutions of the reconnection rate, the current density, and the total resistivity at the X-point are shown in **Figure 1**. The moments of peak reconnection rate $(t_{peak})$ are marked out with red lines, respectively, $t_{peak} = 184\tau_A$ for the $\eta_{spz}$ case and $t_{peak} = 41\tau_A$ for the $\eta_{eff}$ case. With the effective resistivity, the peak reconnection rate $(\sim 0.1B_0 v_A)$ is increased by more than an order of magnitude compared with the $\eta_{spz}$ case $(\sim 0.01B_0 v_A)$. The numerical diffusion is ignorable in the $\eta_{eff}$ case as the $\gamma_{rate}$ calculated by $\partial\psi/\partial t$ (solid line) and $\eta J_y$ (dashed line) in **Figure 1** (d) are almost the same. In contrast, the $\eta_{spz}$ case contains considerable numerical diffusion as indicated by **Figure 1** (a). On the other hand, as the resistivity in the $\eta_{spz}$ case shown in **Figure 1** (c) is a constant, the increase of reconnection rate requires an enhancement of the out-of-plane current density $J_y$ at the X-point [with a peak value of 6.8 compared with the initial value of 2.0, see **Figure 1** (b)]. However, the situation in **Figure 1** (e) is totally different for the $\eta_{eff}$ case, the current density shows a slow increase to the peak value of about 2.3 in the linear stage, and after the start of fast reconnection $(t \approx 23\tau_A)$, $J_y$ decreases quickly to a steady low level $(\approx 1.6)$. As shown by **Figure 1** (f), the total resistivity for the $\eta_{eff}$ case lags a little behind the reconnection rate but exhibits a synergistic growth, and finally reaches a value of 0.08, about a hundred times larger than the Spitzer resistivity.

The 2D distributions of out-of-plane current density with the magnetic field lines at the peak reconnection rate moments are plotted in **Figure 2**, exhibiting significant topological differences for the current sheets. In the $\eta_{spz}$ case, the current sheet is strongly compressed to a long and sharp line with the peak value of 6.8 at the X-point, corresponding to a typical Y-type reconnection. In the $\eta_{eff}$ case, the current sheet width is wider, or almost the same as the initial equilibrium. This is because the resistive dissipation region increases significantly after applying the effective resistivity, as predicted by the Sweet-Parker model $\left(\lambda_b \approx LS_L^{-1/2}\right)$. Besides, the separatrix angle of the $\eta_{eff}$ case is much larger around the X-point but smaller downstream far from the diffusion region if compared with the $\eta_{spz}$ case, that is, the magnetic field topology tends to change from the Y-type into X-type, which in turn increases the effective resistivity around the X-point and further improves the reconnection rate.

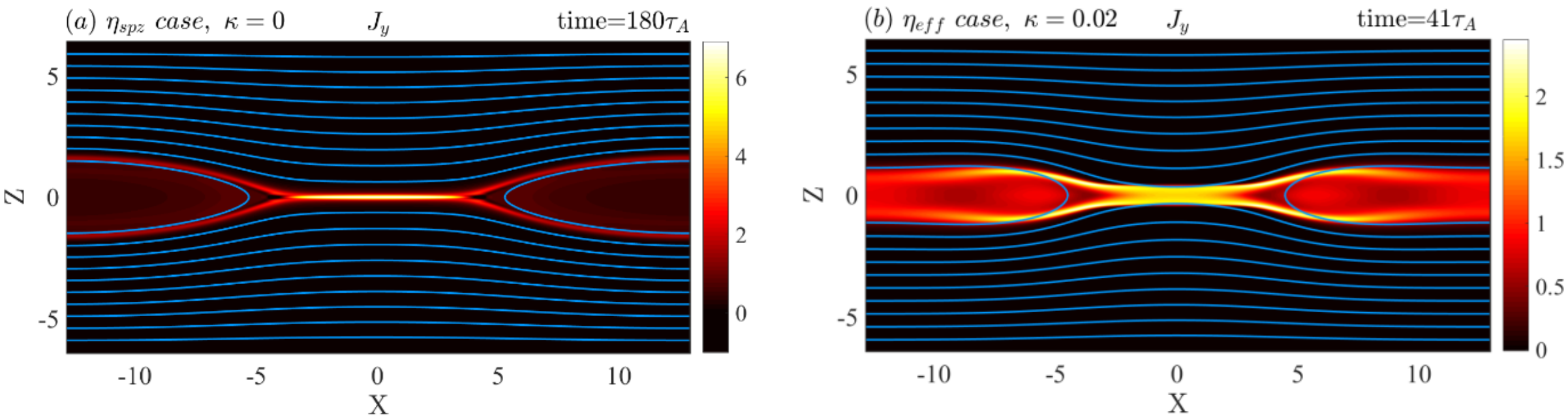

**Figure 2.** (MHD) The distributions of out-of-plane current density with magnetic field lines at the moment of peak reconnection rate, respectively, for (a) the $\eta_{spz}$ case at $t_{peak} = 180\tau_A$ and (b) the $\eta_{eff}$ case at $t_{peak} = 41\tau_A$.

**Figure 3** shows the contribution of each term in Ohm's law to the out-of-plane electric field $[E_y = -(\mathbf{v} \times \mathbf{B})_y + \eta J_y]$ at the neutral line $(z = 0)$ at the peak reconnection rate moment of the $\eta_{eff}$ case. The out-of-plane electric field is mainly sustained by $\eta J_y$ around the X-point where the magnetic field vanishes, while outside the diffusion region, $-(\mathbf{v} \times \mathbf{B})_y$ plays the leading role.

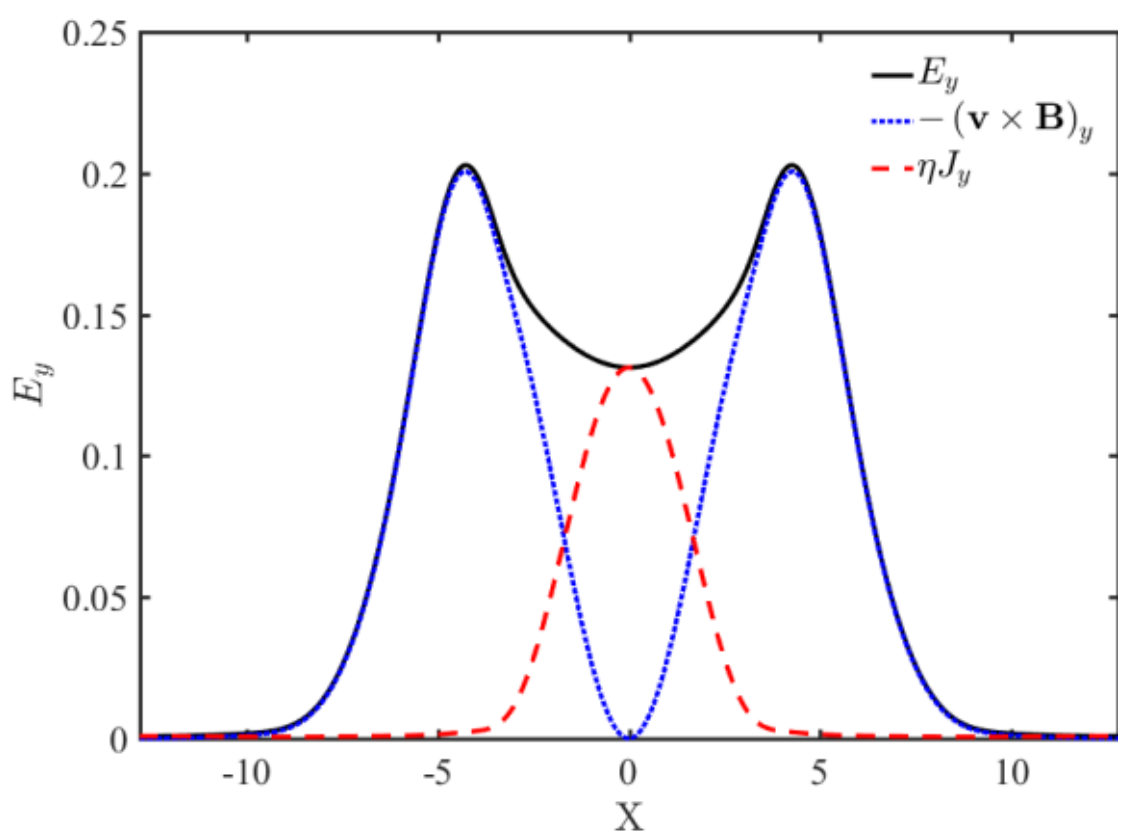

**Figure 3.** (MHD) Contributions in $E_y = -(\mathbf{v}\times\mathbf{B})_y + \eta J_y$ at the neutral line $(z = 0)$ for the out-of-plane electric field at the peak reconnection time $(t_{peak} = 41\tau_A)$ of the $\eta_{eff}$ case.

Through the above comparisons, the traditional resistive MHD model merely considering the low Spitzer resistivity results in the abnormal growth of current density at the X-point and compression on the current sheet, leading to an unexpected reinforcement of the shear field $(B_x \propto J_y\lambda_b)$. However, with the effective resistivity, the reconnection rate depends on more factors as shown by Eq. (10), such as the spatial characteristic of the reconnecting field. Moreover, the effective resistivity generally increases the resistive dissipation, and the enhanced reconnecting field $B_z$ is equivalent to generating a negative $J_y$ in the diffusion region, which is consistent with the decrease of current density at the X-point during fast reconnection.

## 5 Hall MHD simulation results

In this section, we report the simulation results based on the Hall MHD model with and without effective resistivity. All parameters for the $\eta_{spz}$ case $(\kappa = 0)$ and $\eta_{eff}$ case $(\kappa = 0.02)$ with Hall effect are the same as above except that the full Hall term is retained with $d_i = 1.0$.

The time evolutions of the reconnection rate, the current density, and the total resistivity for the $\eta_{spz}$ case and the $\eta_{eff}$ case are shown in **Figure 4**. The employment of effective resistivity in Hall MHD model does not change a lot in the peak reconnection rate calculated by $\partial\psi/\partial t$, both cases show values about $0.25B_0v_A$. However, the reconnection rates calculated by $\eta J_y$ [dashed lines in **Figure 4** (a) and (d)] exhibit significant differences. For the $\eta_{spz}$ case, the peak value of $\partial\psi/\partial t \sim 0.25B_0v_A$ is much larger than $\eta J_y \sim 0.01B_0v_A$. The difference indicates that huge numerical diffusion has been introduced at the X-point for the $\eta_{spz}$ case, which mainly originates from the numerical smoothing performed to stabilize the Hall MHD simulation. However, the situation in $\eta_{eff}$ case is much better. The peak reconnection rates in **Figure 4** (d) represented by $\partial\psi/\partial t \approx 0.25B_0v_A$ and $\eta J_y \approx 0.2B_0v_A$ are comparable with each other. Therefore, the numerical diffusion is significantly reduced after applying the effective resistivity. In addition, the effective resistivity slightly shortens the timescale to reach the peak reconnection rate by about $12\tau_A$. Changing the coefficient $\kappa$ from 0.1 to 0.5 (corresponding to $m_i/m_e$ from 300 to 5000) only modifies the timescale to reach the peak reconnection rate but makes little difference on the peak reconnection rate in Hall MHD simulations with the effective resistivity (not shown), consistent with the previous conclusion that the peak reconnection rate weakly depends on the mass ratio *(Pritchett, 2001*; *Shay et al., 2007*). The evolution of resistivity in the $\eta_{eff}$ case is quantitatively consistent with previous PIC simulation, as shown by Figure 2 (d) in *(Ma et al., 2018*). Specifically, the PIC case with a mass ratio of 400 predicts the peak value of effective resistivity between 0.12 (direct statistical result, i.e., $\eta_{stat} = E_y/J_y$) and 0.18 (based on the effective resistivity model, i.e., $\eta_{eff}$), while the Hall MHD simulation with effective resistivity yield the similar effective resistivity around 0.15 (with the same normalization procedure). The differences in the X-point current density [**Figure 4** (b) and (e)] and the total resistivity [**Figure 4** (c) and (f)] for the $\eta_{spz}$ case and the $\eta_{eff}$ case are similar to the situation of the MHD simulations in Section 4. Therefore, we will not repeat the discussion.

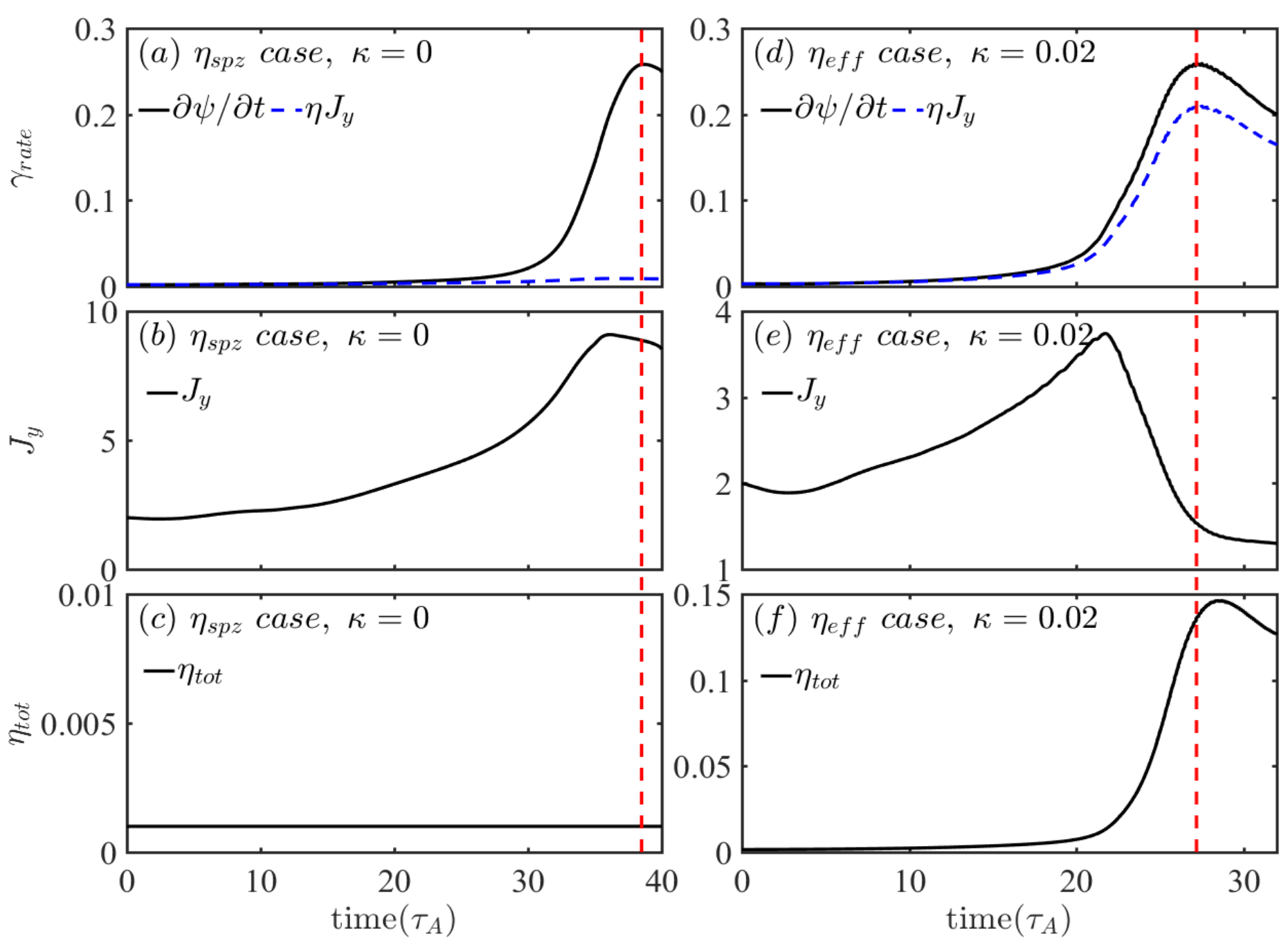

**Figure 4.** (Hall MHD) Time evolution results for (left) the $\eta_{spz}$ case with $\kappa = 0$ and (right) the $\eta_{eff}$ case with $\kappa = 0.02$: (top) reconnection rate calculated by $\partial\psi/\partial t$ (solid line) and $\eta J_y$ (dashed line); (middle) the out-of-plane current density at the X-point; (bottom) the total resistivity. The dashed lines mark the moments of peak reconnection, $t_{peak} = 39\tau_A$ for $\eta_{spz}$ case and $t_{peak} = 27\tau_A$ for $\eta_{eff}$ case, respectively.

**Figure 5** and **Figure 6** show the 2D distributions of the current sheet and quadrupole magnetic field with magnetic field lines at the time of the peak reconnection rate for the $\eta_{spz}$ case and the $\eta_{eff}$ case, respectively. Both cases show obvious X-type magnetic field geometries, consistent with the high reconnection rate of about $0.25B_0 v_A$. Meanwhile, the difference of the quadrupole magnetic field $B_y$ between two cases is negligible, which is mainly determined by the Hall effect outside the diffusion region. Nonetheless, topologies of the current sheets are notably different from each other, a wider current sheet is maintained in the $\eta_{eff}$ case due to larger resistive dissipation, as shown in **Figure 5** (b), while a sharp current singularity forms at the X-point in the $\eta_{spz}$ case in **Figure 5** (a). Besides, the local accumulation of current sheet downstream $(x \approx \pm 5)$ in the $\eta_{eff}$ case is observed, quite similar to the PIC (*Fujimoto and Sydora*, 2008; *Hesse et al.*, 2001a; *Hesse and Winske*, 1998) and hybrid (*Kuznetsova et al.*, 2001) simulation results.

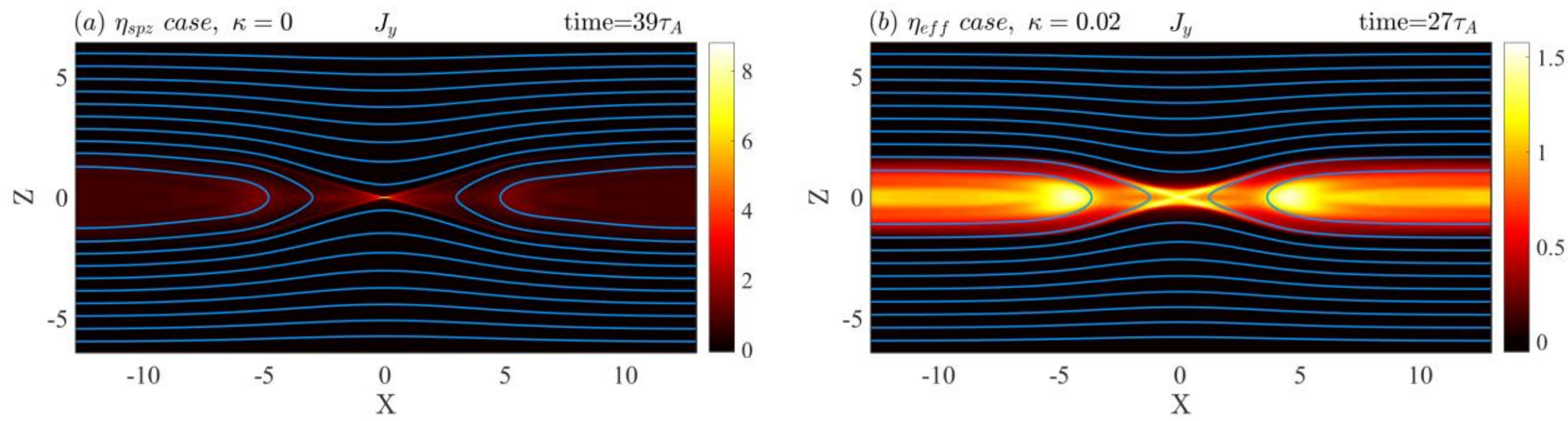

**Figure 5.** (Hall MHD) The distributions of out-of-plane current density with magnetic field lines at the moment of peak reconnection rate, respectively, for (a) the $\eta_{spz}$ case at $t_{peak} = 39\tau_A$ and (b) the $\eta_{eff}$ case at $t_{peak} = 27\tau_A$.

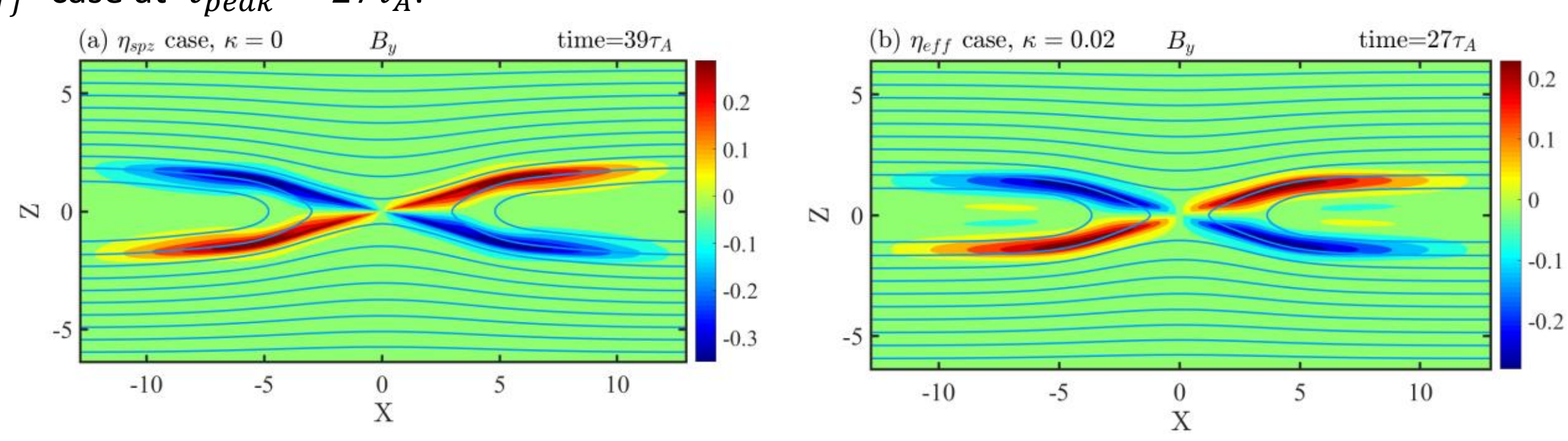

**Figure 6.** (Hall MHD) The distributions of quadrupole magnetic field $B_y$ with magnetic field lines at the moment of peak reconnection rate, respectively, for (a) the $\eta_{spz}$ case at $t_{peak} = 39\tau_A$ and (b) the $\eta_{eff}$ case at $t_{peak} = 27\tau_A$.

We further investigate the different roles played by electrons and ions in the diffusion region. The electron and ion flow velocities are estimated by $\mathbf{v}_e \approx \mathbf{v} - \mathbf{J}/en$ and $\mathbf{v}_i \approx \mathbf{v}$, respectively. Then, the out-of-plane current densities at the X-point for ions and electrons are estimated to be $J_{ey} \approx 0.95J_y$ and $J_{iy} \approx 0.05J_y$. Therefore, the current density and effective resistivity are dominated by electron dynamics at the X-point, consistent with the PIC simulation results with a large ion-electron mass ratio (*Hesse et al.*, 2001a; *Pritchett*, 2001). The in-plane electron and ion flows are plotted in **Figure 7** in the upper and lower half-plane, respectively. In the upstream outside the diffusion region, both electrons and ions are magnetized and moving inward by following the field lines or crossing field due to the $\mathbf{E} \times \mathbf{B}$ drift. However, as closer to the ion diffusion region $(|z| \approx d_i = 1)$, the ions first deviate from the magnetic field lines and are accelerated downstream, while the electrons are still frozen to the field lines and moving much closer to the X-point before leaving the diffusion region, which also explains why the out-of-plane current density at the X-point is dominated by electrons. The electron flows mainly follow the separatrix, leading to the quadrupole field of $B_y$ in Hall MHD or PIC simulations, as shown in **Figure 6**. Inside the diffusion region, where the frozen-in condition $(\mathbf{E} + \mathbf{v}_e \times \mathbf{B} = 0)$ breaks down, due to the bending of the reconnecting field $B_z$, the electrons are scattered away from the X-point with the characteristic time scale of $\tau_e$ of Eq. (5). As a result, the electrons are not allowed undergo continuous out-of-plane electrical acceleration in the diffusion region, which is equivalent to generating an effective resistivity.

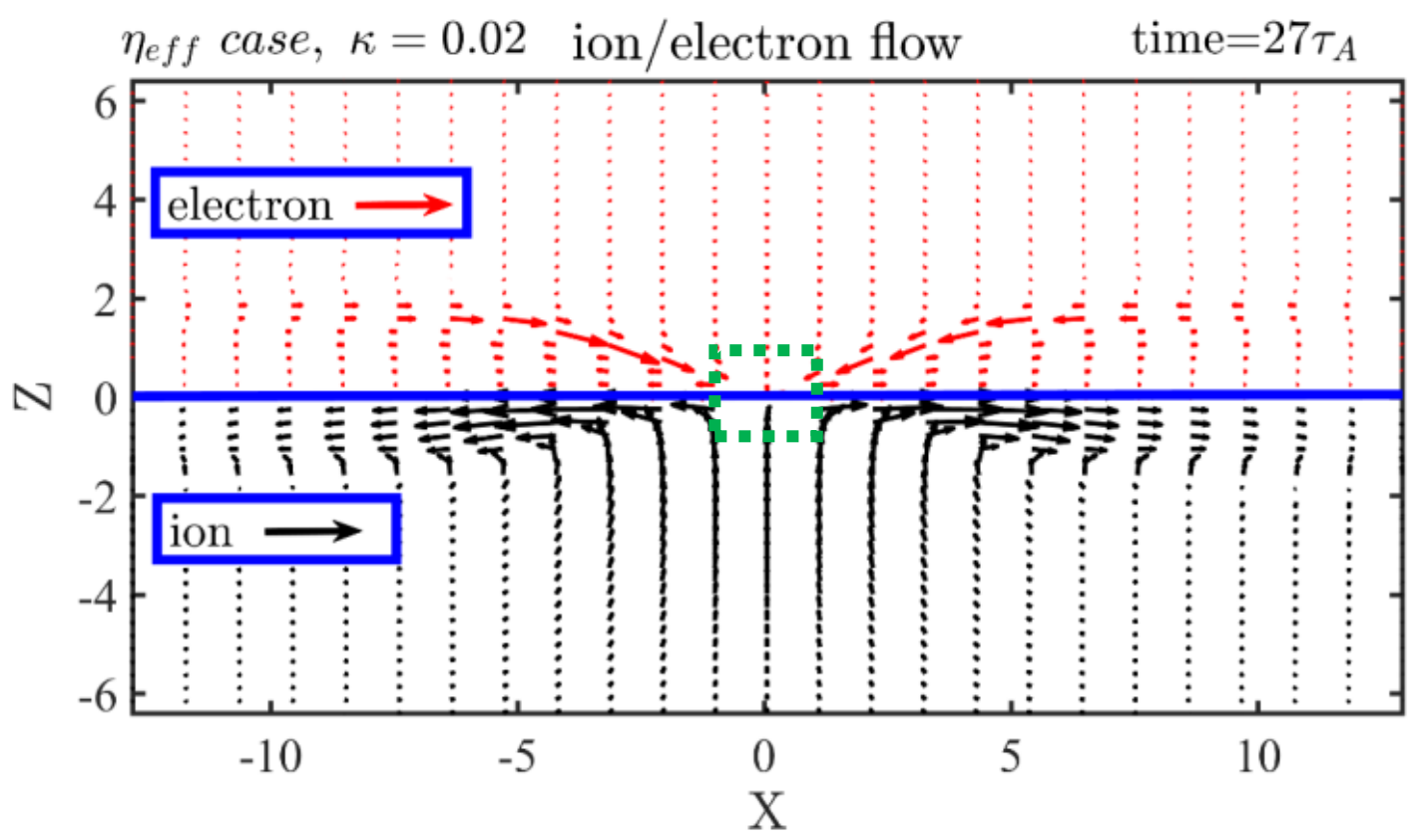

**Figure 7.** (Hall MHD) The in-plane flows of electrons (red, in the upper half plane) and ions (black, in the lower half plane) of the $\eta_{eff}$ case at $t_{peak} = 27\tau_A$. The (ion) diffusion region $(|z| \approx d_i = 1)$ is marked by the green box.

**Figure 8** shows the contribution of each term to the out-of-plane electric field $[E_y = -(\mathbf{v}_e \times \mathbf{B})_y + \eta J_y]$ for the $\eta_{eff}$ case at the neutral line $(z = 0)$ at the moment of peak reconnection rate. The out-of-plane electric field is mainly contributed by $\eta J_y$ at the X-point, while outside the diffusion region, $-(\mathbf{v}_e \times \mathbf{B})_y$ dominates. The discontinuity of $E_y$ in the vicinity of X-point is marked out with an ellipse, where a gap of $\Delta E_y \approx 0.05$ compared with the asymptotic value is seen, close to the difference between the two proxies of the reconnection rate ($\partial\psi/\partial t$ and $\eta J_y$) in **Figure 4** (d). It is mainly caused by the numerical diffusion from the smoothing procedure for the numerical stability in the Hall MHD simulation. That is, a numerical resistivity $\eta_{num}$ should be considered in the diffusion region to satisfy $(\eta_{eff} + \eta_{spz} + \eta_{num})J_y = \partial\psi/\partial t$. It can be expected that in the ideal case without numerical diffusion, the distribution of $E_y$ will be much smoother.

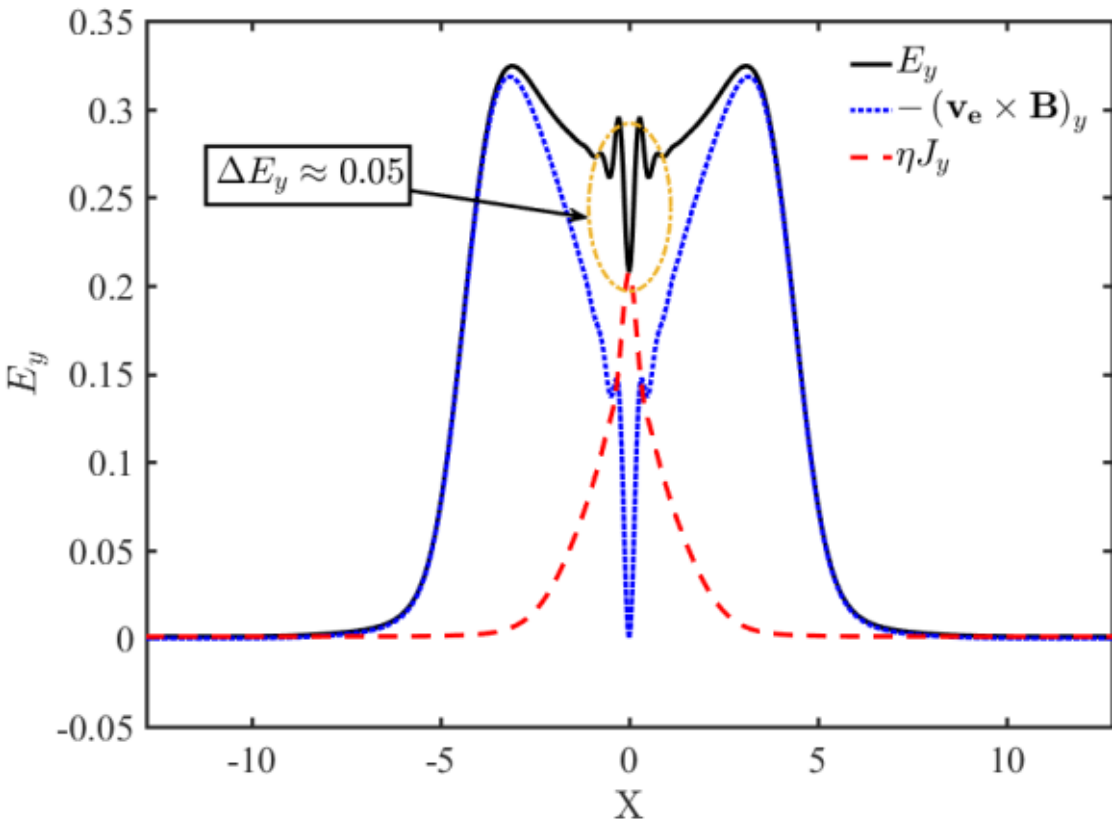

**Figure 8.** (Hall MHD) Contributions in $E_y = -(\mathbf{v}_e \times \mathbf{B})_y + \eta J_y$ at the neutral line $(z = 0)$ for the out-of-plane electric field at the peak reconnection time $(t_{peak} = 27\tau_A)$ of the $\eta_{eff}$ case.

## 6 Conclusion and discussion

The MHD models usually omit the electron inertial term and the anisotropy of electron pressure, thereby losing important electron dynamics during the reconnection process inside the diffusion region. To improve the traditional (Hall) MHD model in collisionless magnetic reconnection simulations, we consider the kinetic features of electrons in a typical reconnecting field and suggest an effective resistivity model *(Ma et al.,* 2018), which is simple and applicable in MHD simulations. The MHD simulations without Hall effect demonstrate the effectiveness of this new resistivity model in speeding up the reconnection process and improving the peak reconnection rate to $\sim 0.1 B_0 v_A$. The topologies of the reconnecting field (tending to be X-type) and current sheet (wider and weaker) are more reasonable compared with the traditional MHD situation. With both the Hall term and effective resistivity, the peak reconnection rate is further enhanced up to $\sim 0.25 B_0 v_A$, close to the PIC results *(Hesse et al.,* 2001a; *Kuznetsova et al.,* 2001). The X-type magnetic geometry, the current sheet splitting, and the ion-electron separation phenomena are consistent with the existing PIC results.

In convectional MHD simulations even with the effective resistivity, the reconnection rate ($\sim 0.1 B_0 v_A$) is still much lower than that of the Hall MHD simulation ($\sim 0.25 B_0 v_A$), which indicates the kinetics of electrons and ions play an equally important role in collisionless magnetic reconnection. In Hall MHD simulations, the breakdown of frozen-in condition of ions just outside the diffusion region roughly determines the peak reconnection rate. Specifically, the Hall term improves the reconnection rate mainly by including the charge separation effect in the presence of Hall current outside the diffusion region, where the electrons are still frozen in the magnetic field lines but the ions become demagnetized. Moreover, the electron dynamics inside the diffusion region are on an equal footing and have been verified by this study and some previous ones, for example, in the form of the off-diagonal electron pressure terms in generalized Ohm's law by PIC simulations (*Cai and Lee*, 1997; *Drake and Shay*, 2007; *Hesse et al.*, 2001b; *Pritchett*, 2001) and laboratory experiments (*Fox et al.*, 2017). Due to the asynchronous demagnetization of different charges, the ions and electrons play the major role respectively in the Hall region and diffusion region. The Hall effect and the effective resistivity work together to enhance the collisionless magnetic reconnection rate. The comparisons for Hall MHD simulation results with and without the effective resistivity indicate the preliminary success of replacing electron kinetic effects in the diffusion region using the analytical effective resistivity model.

The basic theory and the successful application of the effective resistivity model could provide new insights into the anomalous resistivity problem in breaking down the frozen-in condition during the collisionless magnetic reconnection. The influence of the guide field has been preliminarily studied by considering it as a correction to the present effective resistivity model, which requires a more comprehensive calibration with PIC simulations and is beyond the scope of the current paper. Due to the positive feedback nature between the effective resistivity and the reconnection rate, the application of the effective resistivity model in more general magnetic scenarios will require a reliable X-point positioning method to avoid triggering artificial magnetic reconnections. In a practical sense, the effective resistivity model could be applied in global simulations by combining with X-point search methods (*Smiet et al.*, 2020), helping refine the MHD and Hall MHD models, which used to be inadequate for describing the separated ion-electron dynamics, especially in the reconnection diffusion region.

**Acknowledgments**

The author H. W. Zhang would like to acknowledge M. Hoelzl for valuable suggestions and J. S. Wang and J. C. Ren for fruitful discussions. This work is supported by the National MCF Energy R&D Program No. 2022YFE03100001 and 2019YFE03030004.

**Data Availability Statement**

The source data for the MHD simulation results in this paper is available at *Zhang et al.*, 2025.